# Molecular rectifying diodes from self-assembly on silicon

*Stéphane Lenfant , Christophe Krzeminski,*

*Christophe Delerue, Guy Allan & Dominique Vuillaume*[*]

Institute for Electronics, Micro-electronics and Nanotechnology, CNRS,

BP69, Avenue Poincaré, F-59652cedex, Villeneuve d'Ascq, France

ABSTRACT:

We demonstrate a molecular rectifying junction made from a sequential self-assembly on silicon. The device structure consists of only one conjugated ($\pi$) group and an alkyl spacer chain. We obtain rectification ratios up to 37 and threshold voltages for rectification between –0.3V and –0.9V. We show that rectification occurs from resonance through the highest occupied molecular orbital of the $\pi$-group in good agreement with our calculations and internal photoemission spectroscopy. This approach allows us to fabricate molecular rectifying diodes compatible with silicon nanotechnologies for future hybrid circuitries.

---

[*] Corresponding author: vuillaume@isen.iemn.univ-lille1.fr

Since the seminal paper on molecular rectifying diode by Aviram and Ratner[1], molecular electronics has attracted a growing interest with the envisioned perspectives to overcome the foreseen physical limits of conventional microelectronics.[2] Molecular wires based on short oligomers[3-5], carbon nanotubes (CNT)[6-10] and DNA molecules[11-15] were investigated. Hybrid devices (transistors) combining silicon technology and CNT were developed [16-19] Organic monolayers were also used to build molecular switches[20,21] and memory cells.[22,23] Molecular rectifying diodes were synthesized based on the Aviram and Ratner (AR) paradigm,[1] with donor and acceptor moieties linked by a short σ or π bridge.[24-29] This A-b-D molecule is ω-substituted by an alkyl chain to allow a monolayer formation by the Langmuir-Blodgett (LB) method, and this LB monolayer is then sandwiched in a metal/monolayer/metal junction. Up to now, the most interesting results to date were obtained with the hexadecylquinolinium tricyanoquinodimethanide molecule ($C_{16}H_{33}$-Q-3CNQ for short).[24-29] However, the chemical synthesis of this molecule is not obvious with several routes leading to erratic and unreliable results. A more reliable synthesis was reported with a yield of ~59%.[26] Moreover, we have demonstrated[30] that the donor and acceptor groups in this molecule are not sufficiently isolated by the π-bridge, and the molecular orbitals are too delocalized over the entire Q-3CNQ unit to allow a correct implementation of the AR paradigm. The rectification effect was attributed to the geometrical asymmetry induced by the long alkyl chain which places the Q-3CNQ unit close to one of the electrodes (inducing an asymmetric electrostatic profile across the molecule).[30] This asymmetry effect was further theoretically studied more extensively.[31] Based on these considerations, we report an experimental demonstration of a simplified and more robust synthesis of a molecular rectifier with only one donor group and an alkyl spacer chain. We used a sequential self-assembly process (chemisorption directly from solution) on silicon substrates. We analyze the properties of these molecular devices as a function of the alkyl chain length and for two different donor groups.

The chemical synthesis of these self-assembled monolayers (SAM) was described elsewhere[32] and was used previously to fabricate molecular insulator/semiconductor heterostructures.[32] In brief, we started by the self-assembly of vinyl-terminated alkyl chains with several length: Heptadec-16-

en-1-trichlorosilane (HETS, $SiCl_3-(CH_2)_{15}-CH=CH_2$), tetradec-13-en-1-trichlorosilane (TETS, $SiCl_3-(CH_2)_{12}-CH=CH_2$) and oct-7-en-1-trichlorosilane (OETS, $SiCl_3-(CH_2)_6-CH=CH_2$).[33] The substrates were degenerated silicon (resistivity of ~$10^{-3}$ Ω.cm) to avoid any voltage drop in the substrate during electrical measurements. The silicon is covered by its native oxide (1 to 1.5 nm thick as measured by ellipsometry). Extensive wet cleanings (mainly with a piranha solution: $H_2SO_4:H_2O_2$ 2:1, **caution: piranha solution is exothermic and strongly reacts with organics**) and dry cleanings by combining ultraviolet irradiation and ozone atmosphere were performed before starting the self-assembly process. The alkylsilane molecules were dissolved ($10^{-3}$-$10^{-2}$ M) in an organic solvent (70% hexane – 30% carbon tetrachloride) maintained at a constant temperature in a dry nitrogen purged glove box. The freshly cleaned substrate was dipped into the solution for about 1h30. We deposited the HETS at 20°C, the TETS at 6°C and the OETS at -20°C while the corresponding critical temperatures ($T_C$) for optimum deposition (i.e. to form a densely-packed and well-ordered monolayer) are ~26, ~18, and ~-3°C, respectively.[34] The second step consists in modifying the vinyl end-groups by oxidation in aqueous solution of $KMnO_4/NaIO_4/K_2CO_3$ to obtain -COOH terminated monolayers.[35] The yield of this oxidation has been estimated between 70 and 90 %.[36] Finally we grafted conjugated moieties onto the previously formed SAMs using esterification reactions between the -COOH end-groups and two different alcohols (phenylmethanol and thien-3-ylmethanol).[37] Esterifications were carried out during 120h at room temperature in the presence of a water trap (dicyclohexylcarbodiimide, DCCI) to enhance the reaction yield (thien-3-ylmethanol was dissolved in xylene at $10^{-2}$ M, phenylmethanol was used pure). A structural characterization (infrared spectroscopy, ellipsometry and X-ray reflection) of these σ-π SAMs was reported elsewhere.[32,38] Metal (10nm thick aluminium) was evaporated through a shadow mask (electrode area $10^{-4}$ cm$^2$) using an e-beam evaporator. A low deposition rate (~0.1 Å.s$^{-1}$) and a large distance (70 cm) between the sample (hold at a constant temperature of 20°C) and the crucible were used to minor damages on the SAM. More than 20 devices were measured for each combination of alkyl chain lengths and π end-groups. A yield of about 50-70% (ratio of non short-circuited devices over total measured ones) was obtained. Aluminum was chosen to avoid any rectification effect coming from the difference in the work functions of the two electrodes (4.2 eV for Al and 4.1 eV for n$^+$-type Si, electron affinity in that latter case) since it is well known that a largest current is obtained when a positive bias is applied to the electrode with the smallest work function.[39] This effect was observed through metal/SAM/metal junction with Au and Ti electrodes.[40] A schematic representation of these σ-π SAMs is shown in Fig. 1 for a TETS chain.

Figure 1 gives the typical current density vs. voltage (J-V) curves through these σ-π SAMs. The voltage was applied on the Al electrode while the n$^+$-Si substrate was grounded. A rectification effect at negative voltages clearly appears for all of these σ-π SAMs. The rectification ratios, RR=J(at –1V)/J(at 1V), are in the range 2 to 37. This effect is not seen for SAM without the π end-group – inset in Fig. 1 (see also Ref. [41]). Only a very small rectification may be some times observed for the pure alkyl chains but RR is always lower than 5 with statistical variations from sample to sample of the same order of magnitude and with rectification occurring for positive as well as negative voltages (this probably arises from the small difference in the work function between Al and n$^+$-Si and some defects in the SAM). We define a threshold voltage, $V_T$, for rectification as the intercept between a linear extrapolation of the current at high negative voltages and the zero current y-axis (Fig. 1). $V_T$ values are in the range –0.3 to –0.9V with a typical distribution shown in the histograms reported in figure 2. The $V_T$ values are only slightly dependent on the chemical nature of the two end-groups investigated here.[42] We do not observe a significant variation of RR with the chemical nature of the two end-groups nor any correlation with the alkyl spacer length (the sample to sample variations of the RR values are too large – see Fig. 2-b). The performances of these σ-π molecular diodes compete very well with those of more sophisticated molecular diodes previously reported. LB monolayers of $C_{16}H_{33}$-Q-3CNQ molecules[26-29] exhibit average RR of 7.6 (higher value of 27.5) and $V_T$ of the order of 1.6-1.8 V (rectification occurs for positive voltages applied on the electrode at the alkyl side of the molecular junction in that later case, i.e. for negative voltages on the electrode at the π end-group side as in our case, thus $V_T$ values are directly comparable between the two experiments).

We calculated the electronic structure of the silicon/σ-π SAM, combining *ab-initio* and self-consistent tight-binding methods. We consider a model system characterized by a two-dimensional periodicity. It consists of a Si slab containing 9 layers of silicon atoms, which is sufficient to simulate the electronic properties of bulk Si. The σ-π molecules are chemically attached directly on one Si surface (Fig. 3-a) through a Si-C bond. All other dangling bonds at the Si surfaces are passivated by hydrogen. The size of the unit cell is large enough to avoid interaction between the molecules. The electronic structure of the system is calculated using a self-consistent tight binding technique described in Refs. [30,43]. The tight binding parameters for the molecules are given in Ref [43]. For Si, we use the parameters of Ref.[44] that gives a very good band structure in a wide range of energy and in the full Brillouin zone. For Si-C interactions we take the parameters of Robertson.[45] We have not included the Si-O$_2$ layer in the calculations. But we have shown in Ref.[46] that the position of the molecular levels with respect to the Si band structure is only weakly influenced by the SiO$_2$ layer, in agreement with the transitivity rule for band offsets. The calculated density of states (DoS) is given in figure 3-b for the phenyl-terminated monolayers and for two alkyl chain lengths (6 and 12 CH$_2$ groups). A simplified picture of the electronic structure of the Si/SAM/Al junction is also shown (Fig. 3-c). These results show the DoS for the silicon substrate, the π carbon atoms in the aromatic group, the σ carbon atoms in the alkyl chains. The HOMO level (π bonding orbital) lines up with the VB of the silicon, while the LUMO (π* anti-bonding orbital located on the O atom in the ester) is higher in energy, close to the σ* anti-bonding orbital of the alkyl chains. As expected for the chain length investigated here, we did not observed any significant change with the alkyl chain length. Without the π end-group, the current through the Si/SAM/Al structure is due to tunneling through the insulating alkyl monolayers (large σ-σ* gap). With the π end-group, the current is increased by resonant tunneling via the molecular orbitals (MOs) of the π group whose are closer to the Fermi energies of the electrodes than those of the alkyl chain. The rectification effect arises for a negative bias applied on the Al electrode because the energy difference between the Si Fermi energy – $E_{Fsi}$ – (pinned at the conduction band - CB - in the degenerated Si) and the HOMO, $E_{Fsi}-E_H$, is lower than that with the LUMO, $E_L-E_{Fsi}$. If we assume that the π-end group is almost at the Al electrode potential (since this group is at a close contact with the electrode) and that a large part of the potential is dropped in the alkyl chains, a lower voltage (in absolute value) is required to have a resonance between the Si CB and the HOMO when applying a negative bias on the metal electrode than between the Si CB and the LUMO for a positive bias. From our calculations (Fig. 3), $E_0 = E_H - E_{Fsi}$ is about -0.7 eV and -1.1 eV for the thiophene (not shown), and phenyl end-groups, respectively, while the LUMO are located at larger energy (~2.9 eV and ~3.1 eV above $E_{Fsi}$ for the thiophene and phenyl end-groups, respectively). The calculated HOMO-LUMO gaps for these π-end groups attached to the silicon via an alkyl spacer are ~3.6 eV (thiophene) and ~4.2 eV (phenyl). They are smaller than the calculated gaps for the end-groups alone in vacuum (3.8 eV and 4.67 eV for thiophene and phenyl). This reduction may be ascribed to the coupling of the orbitals of the π end-group with the silicon substrate. To determine the experimental position of the HOMO level, the J-V curves were fitted by a one-level model (i.e. a model in which the conduction is dominated by charge transport through a single energy level located at an energy $E_0$ from the electrode Fermi energies). As discussed above, we assume here a resonant effect through the HOMO, then $E_0 = E_H - E_{Fsi}$ ($E_0 < 0$, see Fig. 3-c). The current density is given by,[47,48]

$$J = \frac{2J_0}{\pi} \left\{ \tan^{-1}\left[\theta(|E_0| + \eta eV)\right] - \tan^{-1}\left[\theta(|E_0| - (1-\eta)eV)\right] \right\} \quad (1)$$

where V is the applied potential on the metal electrode (Si is grounded), e is the electron charge, η is the fraction of the potential on the π moiety (i.e. the potential seen by the center of gravity of the π moiety), $J_0$ is the saturation current and θ is an electrode/molecule coupling parameter. The η value was estimated using a simple dielectric model where the σ and π parts of the SAM have a thickness $d_\sigma$ and $d_\pi$ and a dielectric constant $\varepsilon_\sigma$ and $\varepsilon_\pi$, respectively.[27,30]

$$\eta = 1 - \frac{1}{2} \frac{1}{1 + \frac{\varepsilon_\pi d_\sigma}{\varepsilon_\sigma d_\pi}} \quad (2)$$

The thicknesses are known from ellipsometry measurements of the SAM before and after the esterification of the π end-group.[49] Here, the dielectric constants of the thiophene, the phenyl and the alkyl chains are almost similar,[50] so eq. (2) reduces to $\eta=(d_\sigma+0.5d_\pi)/(d_\sigma+d_\pi)$, which simply represents the relative position of the center of gravity of the π moiety measured from the Si substrate. Thus, typical values are $\eta$~0.83, ~0.87 and ~0.9 for the SAMs based on OETS, TETS and HETS, respectively. The measured J-V curves were satisfactorily adjusted with this formula ($R^2 \geq 0.97$). The solid lines (Fig. 1) give typical examples. Figure 4 shows the average $E_0$ experimental values, they are compared with those derived from our calculations. Taking into account the dispersion of the experimental data and the approximations in the calculations (0.2-0.3 eV), we observed a good agreement. Moreover, the calculations were performed for the molecules without the metal electrode on top. In the real device, the π end-groups are in an intimate contact with the Al electrode. The coupling with the metal certainly modifies the position molecular orbitals of the thiophene and phenyl. The same reason may explain that the small difference between the calculated HOMO values for the two π end-groups (c.a. 0.4 eV) is not clearly observed for the $E_0$ values deduced from the experiments.

With the π end-group on top of the alkyl chain, the energy barrier due to the SAM has now a step-like behavior (Fig. 3-c) compare to that of the alkyl chain alone. This step-like shape results from two rectangular barriers in series with energy barrier $\Delta_\sigma$ and $\Delta_\pi$ for the alkyl and aromatic part of the SAM, respectively. In the WKB (Wentzel-Kramers-Brilloin) approximation, these two barriers in series must be considered as an equivalent rectangular barrier (dotted lines in Fig. 3-c) with an average barrier $\Delta$ given by

$$\Delta=\frac{\Delta_\sigma d_\sigma + \Delta_\pi d_\pi}{d} \quad (3)$$

with $d_\sigma$ and $d_\pi$ the thicknesses of the σ and π parts of the SAM ($d=d_\sigma+d_\pi$ is the overall SAM thickness). A significant reduction of the tunneling barrier is thus expected. From previous internal photoemission experiments (IPE) and calculations, the tunneling barriers for a pure alkyl chains were determined (see notation in Fig. 3-c) as $\Delta_{e\sigma}=4.1 \pm 0.2$ eV, $\Delta_{m\sigma}=4.3 \pm 0.2$ eV and $\Delta_{h\sigma}=4.0 \pm 0.2$ eV.[41,46] These values are in agreement with the calculated energy levels for the σ and σ* orbitals shown in Fig. 3-b. Figure 5 shows IPE results for two SAM, the σ-π SAM with a phenyl end-groups and a reference one for the starting SAM (vinyl-terminated alkyl chains). Without the π end-groups, the energy barriers are $\Delta_{e\sigma}$=4.1 ± 0.2 eV, $\Delta_{m\sigma}$=4.2 ± 0.2 eV and $\Delta_{h\sigma}$=4.2 ± 0.2 eV in good agreement with our previous reports (for methyl terminated alkyl chains).[41,46] Clear shifts of the measured photocurrents are observed when adding the π end-group with $\Delta_e$=3.9 ± 0.2 eV, $\Delta_m$=3.9 ± 0.2 eV and $\Delta_h$=3.7 ± 0.2 eV for the phenyl (Fig. 5). Using the known $\Delta_e$ values (quoted above) in eq. (3) and the $\Delta_\pi$ values estimated from the calculated electronic structure (Fig. 3, $\Delta_{e\pi} \sim$ 3.1 eV, $\Delta_{m\pi}$ ~3.2 eV and $\Delta_{h\pi} \sim$ 0 eV), we estimate the following average Δ values: $\Delta_e$~3.9 eV, $\Delta_m$~4 eV and $\Delta_h$~3.5 eV. Owing to the precision limits of the calculations and the error bars in the experiments, these values are quite in agreement with the measured ones (Fig. 5). This feature validates, at first order, the calculated energy position of the HOMO and LUMO of the π end-group relative to the Si CB and Al Fermi energy. In particular, the larger shift for $\Delta_h$ (from 4.2 to 3.7 ev without and with the π end-group, Fig. 5) than for $\Delta_e$ (from 4.1 to 3.9 eV) and $\Delta_m$ (from 4.2 to 3.9 eV) is consistent with the larger energy difference calculated between the σ and π orbitals (see Fig. 3) than between the $\sigma^*$ and $\pi^*$ orbitals (which are almost at the same level, Fig. 3).

Current rectifications were also previously observed for STM tip/vacuum (or air) gap/molecules/metal (or silicon) electrode junctions.[51-53] These features were also ascribed as due to asymmetrical position of the molecules between the electrodes and/or asymmetrical position of the molecular orbitals relative to the energy levels of the electrodes.[54] Moreover, we have previously demonstrated[30] that the rectification in metal/$C_{16}H_{33}$-Q-3CNQ/metal junction is mainly due to a similar geometrical asymmetry of the electrostatic potential induced by the presence of the long alkyl chain rather than by the AR mechanism. A metal/Q-3CNQ/metal junction (molecule with no alkyl chain) would exhibit a symmetrical current-voltage curve.[30] The current rectification by molecules with an asymmetric structure, -$(CH_2)_m$-$C_6H_6$-$(CH_2)_n$- with m=2 and n=2 to 10, was also recently calculated leading to a similar conclusion that the rectification ratio increases with the ratio n/m.[31] In the present work, we exploited and controlled, using a sequential assembly, the geometrical molecular asymmetry in a σ-π SAM to design a molecular diode. We clearly demonstrated that the current asymmetry comes from a resonant tunneling through the HOMO level of the π-end group.

Rectification ratios (RR) are in the range 2 to 37 and threshold voltages for rectification are in the range −0.3 to −0.9 V. However we did not observed the predicted trends (RR increases with the alkyl chain length and the absolute value of $V_T$ decreases with the alkyl chain length).[31] This is due to the relatively large sample-to-sample variations of the experimental data. Moreover, our device structure is not exactly similar to the model structure used in Ref. [31], the small tunnel barrier (2 $CH_2$ groups) is not present here. We may just surmise that an ultra-thin aluminum oxide may form at the SAM/Al interface when the samples are brought to ambient air after the electrode evaporation. Further works are in progress to reduce the experimental dispersion and to study other π end-groups. Changing the energy location of their molecular orbitals would allow us to tune the rectification properties (e.g. $V_T$). In conclusion, the proposed approach allows building a molecular rectifying diode compatible with silicon nanotechnologies for future hybrid circuitries with simple guidelines to tune its rectification performances.

ACKNOWLEDGEMENTS: Work financially supported by the French Ministry of Research (ACI Nanostructures). We thank O. Bouloussa and S. Palacin for the synthesis of the TETS and HETS molecules, respectively, and J. Collet for some of the IPE measurements.

(42) However, other end-groups with quite different positions of their molecular orbitals with respect to the electronic structure of the electrodes can probably be used, which can induce a significant change in the threshold voltages.

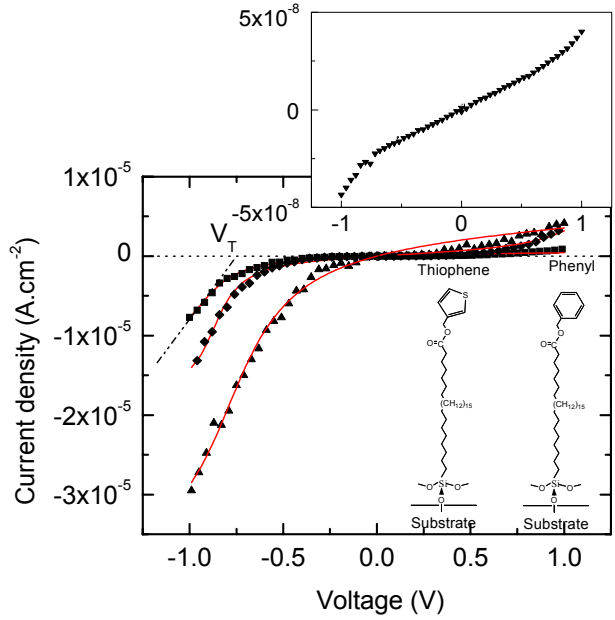

**Figure 1**. Figure caption.   Current-voltage characteristics of the Si/σ-π SAM/Al junctions: (▲) OETS thiophene-terminated SAM; (■) HETS phenyl-terminated SAM; (♦) HETS thiophene-terminated SAM. All measurements were taken at room temperature in the dark. We used a shielded probe station equipped with micromanipulators. The probe station was purged by a continuous flow of dry nitrogen. The inset shows the reference curve for the starting HETS SAM (i.e. without the π end-group). The solid lines are the fitted curves of eq. (1) with the following parameters: $J_0 = 1.57 \times 10^{-5}$ A.cm$^{-2}$, $\theta = 4$ eV$^{-1}$, $E_0 = -0.67$ eV, $\eta \approx 0.83$ for the OETS-thiophene SAM; $J_0 = 6.59 \times 10^{-6}$ A.cm$^{-2}$, $\theta = 7.4$ eV$^{-1}$, $E_0 = -0.84$ eV, $\eta \approx 0.9$ for the HETS-phenyl SAM; and $J_0 = 1.57 \times 10^{-5}$ A.cm$^{-2}$, $\theta = 7.0$ eV$^{-1}$, $E_0 = -0.88$ eV, $\eta \approx 0.9$ for the HETS-thiophene SAM. $\eta$ is not a fitted parameter, its value is estimated from thickness measurements (see text). The graphical determination of the threshold voltage for rectification ($V_T$) is also shown for one case. $V_T$ is taken as the intercept between the linear extrapolation of the J-V curve and the zero of the y-axis. The typical $V_T$ value is -0.71V for this HETS-phenyl SAM. Rectification ratios are 8.5, 5 and 7.1 for these HETS-phenyl, HETS-thiophene and OETS-thiophene devices.

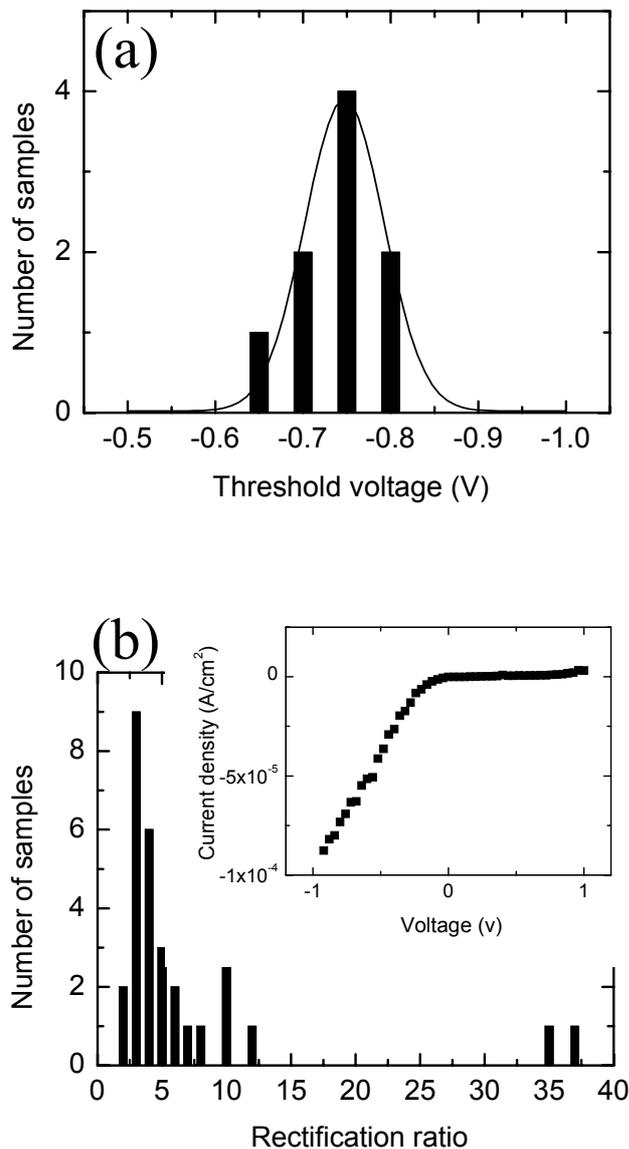

**Figure 2**. Figure caption      (a) Typical histogram of the threshold voltage for rectification (TETS-thiophene diode). The solid line is the fit by a Gaussian function given an averaged value of $V_T$=-0.74V. The other averaged values (not shown) are –0.35V and –0.72V for the OETS- and HETS-thiophene diodes, and –0.42V, -0.68V and –0.73V for the OETS-, TETS- and HETS-phenyl diodes, respectively. (b) Histogram of the rectification ratio for all the measured devices. The inset shows a typical J-V curve for the device with the best RR (OETS-thiophene).

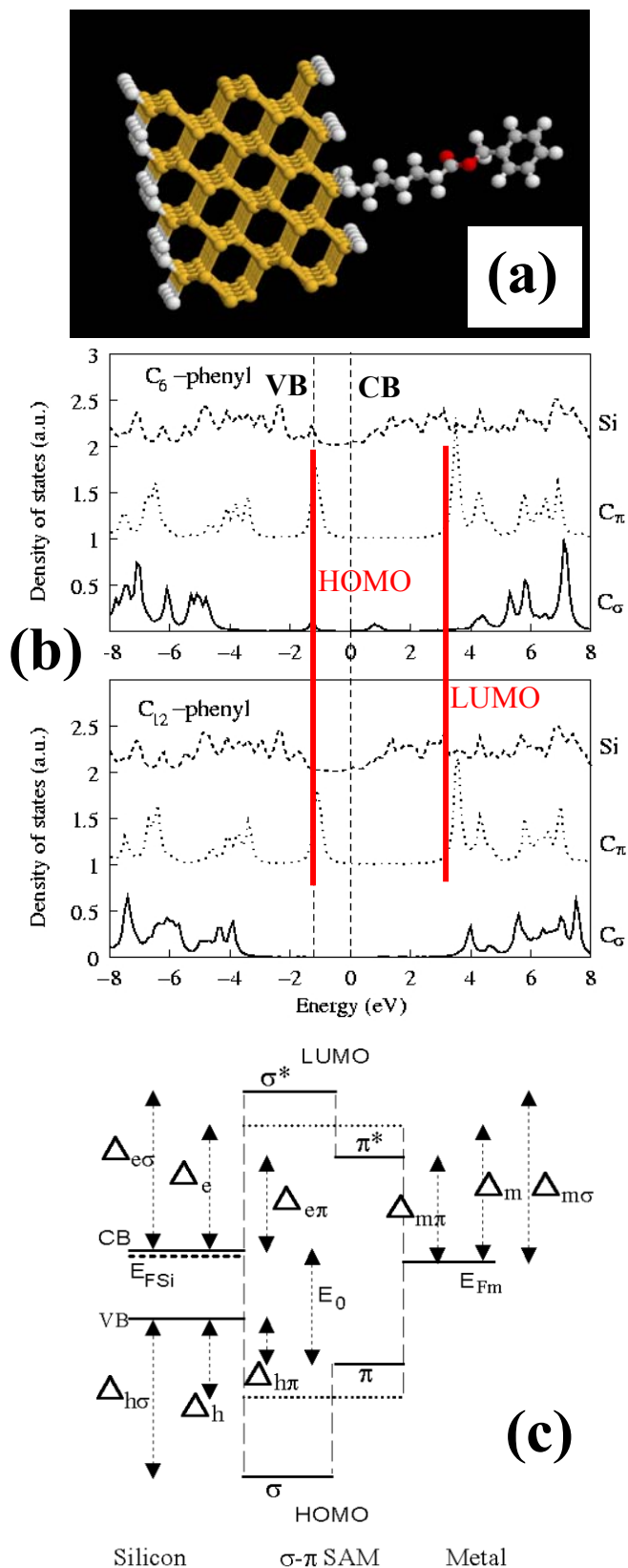

**Figure 3**. Figure caption (a) Ball and stick representation of the model system used for the calculation: yellow=Si, white=H, grey=C, red=O (b) Calculated local density of states (DoS) for the π carbon atoms, σ carbon atoms and silicon substrate. The calculation was done for the case of TETS and OETS alkyl chains with a thiophene or a phenyl end-group (only the results for the phenyl

are shown). The zero energy (reference) is at the silicon conduction band (i.e. also the Fermi level in the n-type highly-doped Si). The position of the silicon VB is positioned at 1.1 eV below the energy reference. The HOMO and LUMO of the π end-groups are also indicated (red lines, the LUMO is located on the O atom of the ester, not shown). For the phenyl, we have the HOMO at –1.1 eV below the Si CB ($E_0$ in Fig. 3-c) and the LUMO at +3.1 eV above it ($\Delta_{e\pi}$ in Fig. 3-c). For the thiophene (not shown), the HOMO is at –0.7eV and the LUMO at +2.9eV. (c) Schematic energy diagram of the Si/σ-π SAM/Al junction at zero bias determined from these calculations. The energy barrier between the two electrodes is taken as a step-like rectangular barrier with energy barriers deduced from the calculated DoS (Fig. 3-b): $\Delta_{e\pi}$=2.9 eV, $\Delta_{m\pi}$=3 eV and $\Delta_{h\pi}$= -0.4 eV for the thiophene end-group; $\Delta_{e\pi}$=3.1 eV, $\Delta_{m\pi}$=3.2 eV and $\Delta_{h\pi}$= 0 eV for the phenyl end-group. For the alkyl chain, we have $\Delta_{e\sigma}$=4.10 eV, $\Delta_{m\sigma}$=4.20 eV and $\Delta_{h\sigma}$=4.20 eV (see details of calculations in [46]). Note that the energy barrier for electron at the metal/SAM electrode was simply deduced from the one at the silicon/SAM +0.1 eV, 0.1 eV being the work function difference between the Al electrode and the highly-doped silicon substrate (4.2 eV for Al and 4.1 eV for n$^+$-type Si, electron affinity in that latter case). The dotted lines are the average rectangular energy barrier as determined by IPE experiments (see text, eq. 3 and Fig. 5) with average energy barriers $\Delta_e$, $\Delta_h$ and $\Delta_m$.

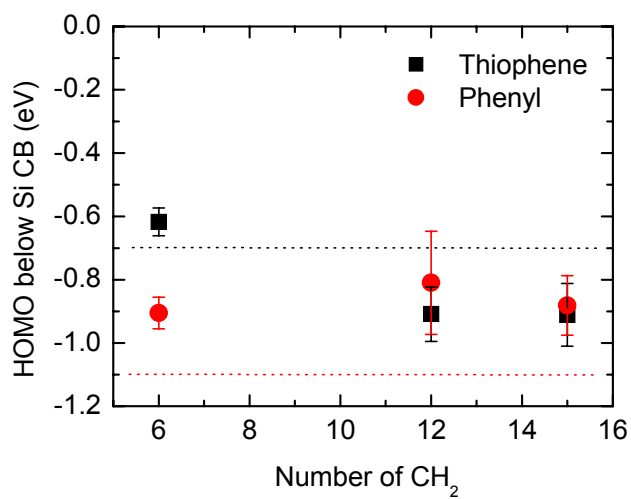

**Figure 4**. Figure caption    Energy level ($E_0$) through which the charge transport occurs at the rectification (negative) potential. $E_0$ was deduced from the fit of Eq. (1). The plot gives the averaged values, the error bars stand for the standard deviation of the distribution observed on a set of measurements over about 5-10 samples for each combination of alkyl chain length and π end-group. The dotted lines represent the calculated $E_0$ values for the thiophene (at –0.7 eV) and for the phenyl (at –1.1 eV) end-groups. Zero energy is the Fermi energy in the silicon (pinned at the CB for the n-type highly-doped substrate used in this work).

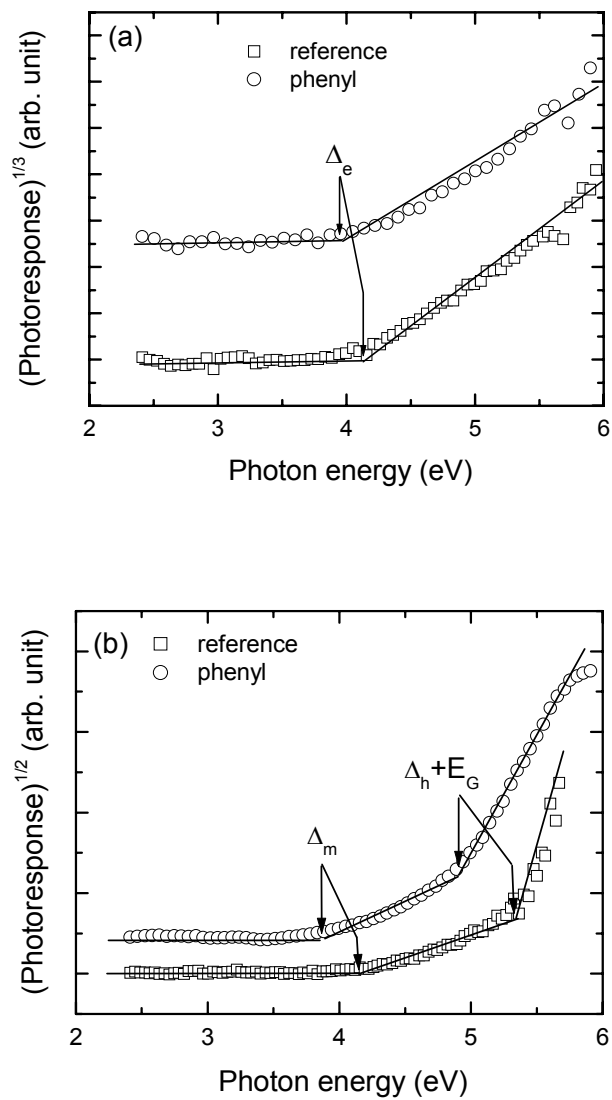

**Figure 5**. Figure caption    Photocurrent response versus the photon energy for a positive bias (+100mV) on the Al electrode (a) and for a negative (-100mV) bias on the Al electrode (b). For a positive bias, electrons are injected from the Si CB over the LUMO of the SAM thus the threshold is a measure of the barrier height Δe. Above the photoinjection threshold, linear variations are expected when plotting the cube root of the photoresponse (photocurrent normalized to the incident photon flux). For a negative bias, the photocurrent is due to electrons injected from the metal Fermi level (energy barrier Δm) and a second contribution is due to holes injected from empty states of the Si substrate for photon energies higher than Δh+Eg (Eg being the silicon band-gap=1.1 eV). In this case, above threshold a linear variation is observed when plotting the square root of the photoresponse (see details of the experiments in 41,46,55). The different curves are shifted along the y-axis for clarity. The values of Δe, Δm and Δh are extrapolated from the linear parts of the measured curves (straight lines), the horizontal lines are the zero photocurrent level. We have Δeσ = 4.1 ± 0.2 eV, Δmσ = 4.2 ± 0.2 eV and Δhσ = 4.2 ± 0.2 eV for the TETS SAM (no π end-group) and Δe = 3.9 ± 0.2 eV, Δm = 3.9 ± 0.2 eV and Δh = 3.7 ± 0.2 eV for the TETS-phenyl SAM.

ToC graphic: 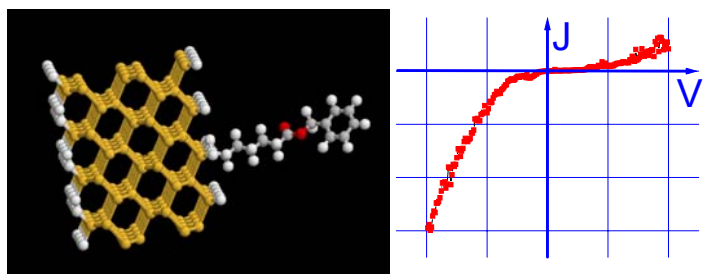